\documentclass[%
 reprint,
superscriptaddress,
 amsmath,amssymb,
 aps,
 prl
]{revtex4-2}

\usepackage{dsfont}
\usepackage{graphicx}
\usepackage{dcolumn}
\usepackage{bm}
\usepackage{hyperref}
\hypersetup{
colorlinks=true,
	linkcolor=nrppurple,
	citecolor=nrppurple,
	urlcolor=nrppurple,
}

\usepackage{amssymb}
\usepackage{amsmath}
\usepackage{siunitx}
\usepackage{xcolor}
\usepackage{dsfont}	
\usepackage{mathrsfs}	

\renewcommand{\v}[1]{\bm{\mathrm{#1}}}

\newcommand{\p}{\v{p}}

\newcommand{\df}{\mathrm{d}}

\newcommand{\tr}{\mathrm{tr}}

\newcommand{\Tr}[0]{\mathbf{Tr}}	

\newcommand{\w}{\omega}

\makeatletter \renewcommand\@make@capt@title[2]{%
\@ifx@empty\float@link{\@firstofone}{\expandafter\href\expandafter{\float@link}}%
\sffamily{\textbf{#1}}\@caption@fignum@sep#2 }
\usepackage{placeins} 

\begin{document}
\definecolor{nrppurple}{RGB}{128,0,128}

\preprint{APS/123-QED}

\title{Proximity-induced collective modes in an unconventional superconductor heterostructure}
\author{Jonathan B. Curtis}
\affiliation{John A. Paulson School of Applied Sciences and Engineering, Harvard University, Cambridge Massachusetts 02138 USA}
\affiliation{Department of Physics, Harvard University, Cambridge, MA 02138, USA}
\author{Nicholas R. Poniatowski}
\affiliation{Department of Physics, Harvard University, Cambridge, MA 02138, USA}
\author{Amir Yacoby}
\affiliation{John A. Paulson School of Applied Sciences and Engineering, Harvard University, Cambridge Massachusetts 02138 USA}
\affiliation{Department of Physics, Harvard University, Cambridge, MA 02138, USA}
\author{Prineha Narang}
\email[]{prineha@seas.harvard.edu}
\affiliation{John A. Paulson School of Applied Sciences and Engineering, Harvard University, Cambridge Massachusetts 02138 USA}

\date{\today}
 
\begin{abstract} 
Unconventional superconductors have been long sought for their potential applications in quantum technologies and devices.
A key challenge impeding this effort is the difficulty associated with probing and characterizing candidate materials and establishing their order parameter.
In this \emph{Letter}, we present a platform that allows us to spectroscopically probe unconventional superconductivity in thin-layer materials \emph{via} the proximity effect.
We show that inducing an $s$-wave gap in a sample with an intrinsic $d$-wave instability leads to the formation of bound-states of quasiparticle pairs, which manifest as a collective mode in the $d$-wave channel.
This finding provides a way to study the underlying pairing interactions vicariously through the collective mode spectrum of the system. 
Upon further cooling of the system we observe that this mode softens considerably and may even condense, signaling the onset of time-reversal symmetry breaking superconductivity.
Therefore, our proposal also allows for the creation and study of these elusive unconventional states.
\end{abstract}

\maketitle

Materials exhibiting unconventional superconductivity are key components of many proposed quantum devices.
For instance, triplet superconductors may allow for the incorporation of magnetic functionalities into superconducting electronics, as well as offering larger critical magnetic field strengths~\cite{Linder.2015,Yang2021}.
Similarly, a large amount of work has been devoted towards realizing topological superconductors, such as the elusive chiral $p$-wave state, due to their potential for realizing topologically protected Majorana-based qubits and quantum computation~\cite{Alicea_review,sankar-rmp,Zareapour.2012}.
Many of these useful unconventional states break additional symmetries, beyond global $U(1)$ symmetry, such as time-reversal symmetry~\cite{Sigrist2000,hirschfeld-s+id,mixedsymm,valentin-1,trsb-review,Wysokinski.2019}. 

The question of how to realize~\cite{Narang.2020,Zareapour.2012} and prepare these systems not withstanding, it is often very difficult to even characterize and verify the nature of these unconventional superconducting phases.
Often, low dimensionality, low temperature scales, and complex order parameters can conspire to obscure the microscopic structure of the ground state, making the unambiguous identification of the state challenging. It has recently been emphasized that one potential solution to this problem is to use the spectrum of collective modes in the superconductor to look for signatures of the ground state order \cite{higgs-spect,higgs-spect2,poniatowski2021spectroscopic}. 
For example, one may look at the multiple different Higgs modes of an anisotropic superconductor in order to identify the ground state symmetry~\cite{higgs-spect,nbn-thg,chiral-higgs,Nosarzeski.2017}.
Similarly, in the case of multicomponent~\cite{poniatowski2021spectroscopic,214-clapping,Balatsky.2000,214-clapping} or multiband~\cite{multiband-review,Marciani.2013,valentin-1} superconductors which break time-reversal symmetry, it has been argued that collective modes associated to the relative phase stiffnesses may also provide signatures of the time-reversal symmetry breaking.

While promising, this method is greatly restricted in its applicability.
In order to support these collective modes, the material must already have two or more closely competing interactions, and if the system has a nodal order parameter there is an additional threat due to quasiparticle damping.
In addition, the relevant frequency scales for these collective modes is almost always on the order of the electronic gap, and therefore usually falls within a challenging frequency range of low-to-mid THz.

In this \emph{Letter}, we present a way to overcome these challenges in a controlled and tunable manner by using the proximity effect to build a ``designer" collective mode.
This collective mode can then be used to probe the order parameter of a candidate material by standard means such as Raman or tunneling spectroscopy. 
Additionally, we show that this protocol may also yield a way to engineer systems which spontaneously breaks time-reversal symmetry, offering a way to systematically study these elusive superconducting states. 

\begin{figure}
    \centering
    \includegraphics[width=\linewidth]{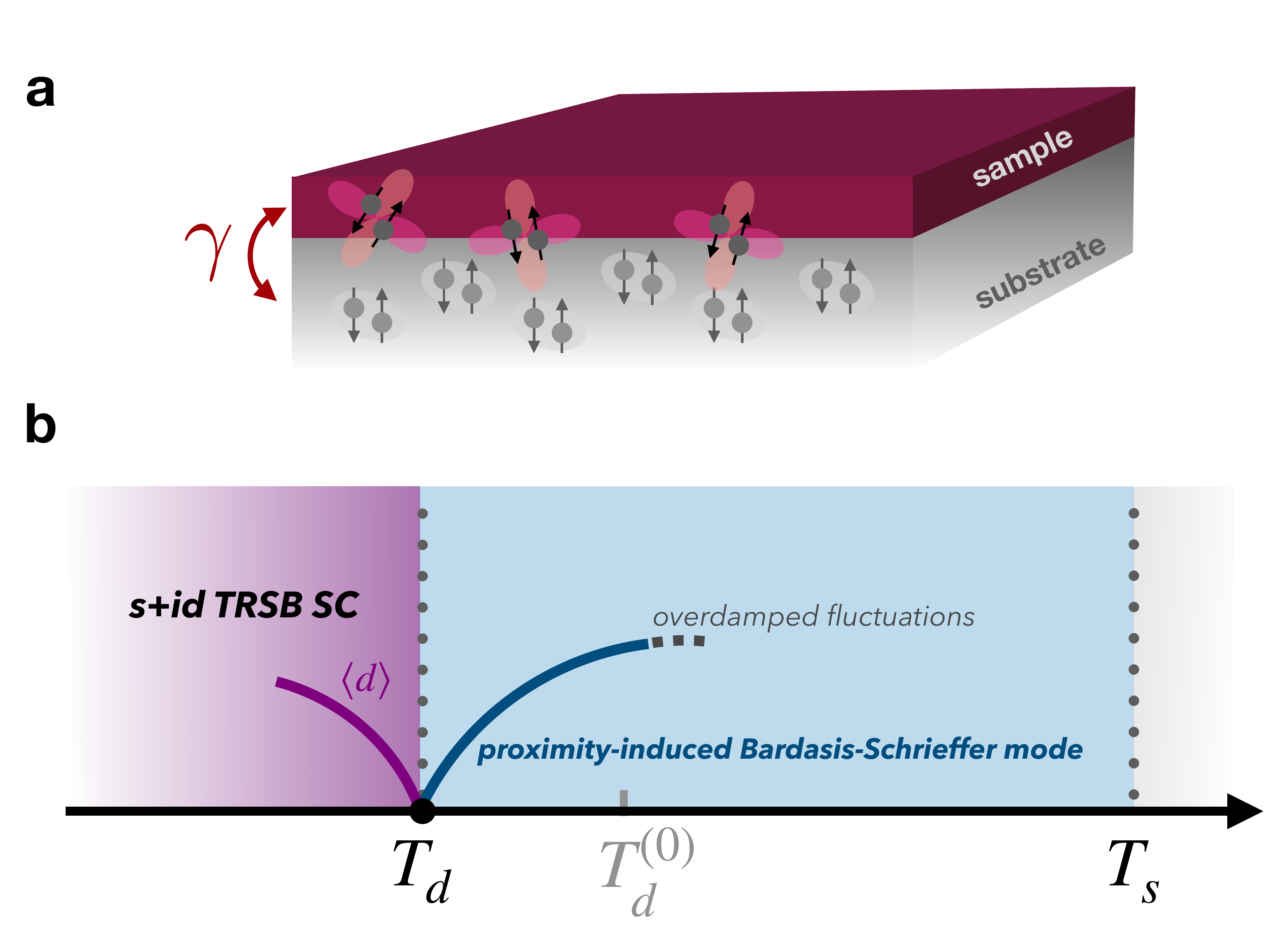}
    \caption{\textbf{a.} Schematic of the heterostructure under study in this work: an unconventional superconducting sample placed in proximity to a conventional bulk superconducting substrate. The two sub-systems are coupled via single-particle tunneling which occurs at the rate $\gamma$. \textbf{b.} Phase diagram of the system. Below $T_s$, the substrate is superconducting and induces a minigap in the sample via the proximity effect. When this minigap becomes large enough, it converts the overdamped fluctuations of the $d$-wave superconducting order into a sharp collective mode. As the sample transition temperature $T_d$ is approached, this proximitized collective mode softens and ultimately condenses out-of-phase with the substrate order parameter, spontaneously breaking time-reversal symmetry. Below $T_d$, the proximity-induced collective mode becomes the usual clapping mode in time-reversal symmetry breaking superconductors. }
    \label{fig:overview}
\end{figure}

Fundamentally, our scheme relies on using a bulk ``substrate" $s$-wave superconductor to proximity induce $s$-wave superconductivity in a thin ``sample" layer of unconventional superconductor which has an intrinsic instability towards pairing in a non-$s$-wave channel, as depicted in Fig.~\ref{fig:overview}(a).
In the presence of the proximity-induced minigap this residual interaction manifests through the formation of stable bound-states of pairs of quasiparticles. 
These bound states essentially realize the Bardasis-Schrieffer, or ``particle-particle exciton," collective mode~\cite{bs-modes}, but in this case the sub-dominant pairing interaction is the dominant pairing interaction in the sample. 

More generally, this realizes a scenario where the substrate and sample order parameters are ``competing"~\cite{hirschfeld-s+id,mixedsymm}.
If the intrinsic pairing interaction in the sample is sufficiently strong then the particle-particle bound-state may itself condense, at which point we show the system will undergo a second phase transition into a state with mixed order in the two channels, and that this will spontaneously break time-reversal symmetry. 
This hierarchy of temperatures and the various regimes are shown in Fig.~\ref{fig:overview}(b).

For specificity, we will demonstrate this idea by considering a concrete model where the sample has single-band spin-singlet $d_{x^2 - y^2}$-wave order, though many of our results are expected to generalize to more complex sample order parameters. 
For simplicity, we take the sample thickness to be thin compared to the coherence length in the out-of-plane direction, such that we may neglect the dispersion in this direction, and hence neglect the spatial dependence of the problem in the transverse direction.

In this case, we describe the intrinsic pairing interaction in the sample by a BCS Hamiltonian 
\begin{multline}
\label{eqn:sample-H}
    H_{0} = \sum_{\bf p \sigma } \xi_{\bf p} c_{\bf p\sigma}^\dagger c_{\bf p\sigma } \\
    - g_d \sum_{\bf q}\int_{\bf p,p'} \chi^d_{\bf p}\chi^d_{\bf p'} c_{\mathbf{p}'+\frac12\mathbf{q},\uparrow }^\dagger c_{-\mathbf{p}'+\frac12\mathbf{q},\downarrow }^\dagger c_{-\mathbf{p}+\frac12\mathbf{q}\downarrow} c_{\mathbf{p}+\frac12\mathbf{q} \uparrow} ,
\end{multline}
with the dispersion relation $\xi_{\bf p} = \mathbf{p}^2 /2m - E_F$ and $d$-wave form-factor $\chi^d_{\bf p} = \sqrt{2}\cos(2\theta_{\bf p})$ (the momentum angle is measured from the $x$-axis, ultimately set by the crystal lattice). 
The second term describes the scattering of Cooper pairs with center-of-mass momentum $\bf q$ and relative momentum $\bf p$, and in particular projects the interaction onto the $d$-wave singlet channel.

Within mean-field theory the pairing interaction can be decoupled, yielding the standard Bogoliubov-de Gennes Hamiltonian for quasiparticles. 
Solving this self-consistently for the $d$-wave gap 
\begin{equation}
\label{eqn:d-wave-gap}
    \frac{1}{g_{d}} \Delta^d = \int_{\bf p} \chi^d_{\bf p} \langle c_{-\bf p \downarrow} c_{\bf p \uparrow} \rangle ,
\end{equation}
we find that $d$-wave pairing sets in at a temperature $T_d^{(0)}$ which is given by the standard BCS formula in terms of the dimensionless pairing strength $\lambda_d = g_d \nu_F $ (with $\nu_F$ the density-of-states at the Fermi level), and a UV cutoff $\Lambda$ of order of the characteristic frequency of whatever mediates pairing in the sample (for phonons this is the standard Debye frequency). 
In this work we do not consider any changes to the intrinsic interaction due to the substrate, though relaxing this is an interesting direction for future study. 

We now introduce the coupling to the substrate, which we treat as a fixed ``reservoir," that doesn't experience any back-reaction due to the coupling to the sample. 
In particular, we assume the substrate is much thicker than the sample and the $s$-wave coherence length.
Crucially, we also assume that the substrate transition temperature $T_s$ is much larger than the intrinsic transition temperature in the sample $T_d^{(0)}$, or equivalently that the substrate superconducting gap $|\Delta_s|$ has largely saturated once the temperature reaches $T \sim T_d^{(0)}$. 

We assume a local tunneling in to the substrate with effective tunneling matrix element $\mathfrak{t}$.
At second-order we find the tunneling energy scale $\gamma = 2\pi \nu_s |\mathfrak{t}|^2$, where $\nu_s$ is the density-of-states in the substrate (for details see Supplemental Material). 
We largely focus on the regime $\Delta_s \gg \gamma$, such that the tunneling scale is less than the substrate gap (for a less restrictive treatment, see Supplemental Material) and we may treat processes only in the Andreev channel.

Provided the substrate superconducting phase is not strongly fluctuating (the relevant energy and length scales over which the phase varies are the plasma frequency and the in-plane penetration depth of the substrate, respectively), we can model the proximity-induced superconducting gap in the sample by adding a term to the Hamiltonian (see Supplemental Material for derivation),
\begin{equation}
\label{eqn:prox}
    H_{\rm prox} = -\frac12 \gamma \sum_{\bf p}  c_{-\bf p\downarrow } c_{\bf p \uparrow } \frac{\overline{\Delta_s} }{|\Delta_s|} + c_{\bf p\uparrow }^\dagger c_{-\bf p \downarrow }^\dagger \frac{\Delta_s}{|\Delta_s|}. 
\end{equation}
In particular, this opens a minigap at the Fermi level for the electrons in the sample (in the Andreev regime $|\Delta_s| \gg \gamma$, the size of the minigap is $\gamma/2$) and the phase is referenced with respect to the substrate phase.
We henceforth set this phase to be zero, such that the substrate gap is taken to be real and positive. 

Since the proximity effect opens a gap on the Fermi surface of the sample, we expect the intrinsic $d$-wave pairing transition to be suppressed. 
Indeed, by solving the mean-field equations in the $d$-wave channel and incorporating the new proximity gap, we find that there is a depression of the critical temperature to $T_d < T_d^{(0)}$.
These results are summarized in Fig.~\ref{fig:tc}a, where we use the frequency-dependent Matsubara Green's function, derived in the Supplemental Material, to solve Eq.~\eqref{eqn:d-wave-gap} and identify the critical temperature in the $d$-wave channel. 

We now proceed to our main result, which is the emergence of the bound-state collective mode. 
Above the new $d$-wave transition temperature, the $d$-wave order is uncondensed but still fluctuates due to the remnant pairing interaction.
Within the Random Phase Approximation we may derive an equation of motion which describes the dynamics of this fluctuating $d$-wave order.
This is derived in detail in the Supplemental Material, but it may be understood as the linear-response pair-susceptibility of the sample in the proximitized state~\cite{Scalapino.1970}.
The presence of a bound-state collective mode then shows up as a resonance in the pair-susceptibility. 

We separate the $d$-wave order parameter into the components which are in-phase and out-of-phase with respect to the substrate condensate, writing $\Delta^d_{\bf q}(t) = h_{\bf q}(t) + i d_{\bf q}(t)$.
We find that the in-phase component $h_{\bf q}$ has no sharp resonance and essentially mirrors the two-particle continuum, and thus we will henceforth neglect the in-phase component.
This is in line with the expectation that the $s$- and $d$-wave orders are competing and therefore the ``repulsion" between the two orders is minimized when they are mutually out of phase~\cite{mixedsymm}.

At linear order we calculate the spectral function for the dynamic pair-susceptibility in the out-of-phase fluctuation $d_{\bf q}$ channel
\begin{equation}
   \mathcal{A}_{dd}(\Omega,{\bf q}) = -\frac{1}{\pi}\Im\left\{ -i \int_0^{\infty}dt e^{i\Omega t} \langle [ d_{\bf q}(t), d_{-\bf q}(0)]\rangle \right\} , 
\end{equation}
which in particular captures the binding energy and linewidth of the $d$-wave excitation.
The spectral function $\mathcal{A}_{dd}(\Omega,{\bf q})$ is obtained in the Supplemental Material in terms of the Nambu-Gor'kov Green's functions using the Keldysh technique, although it may also be calculated using, e.g. the Anderson pseudo-spin method provided $|\Delta_s | \gg \gamma$, so that retardation and damping due to the substrate may be safely neglected.

\begin{figure}
    \centering
    \includegraphics[width=\linewidth]{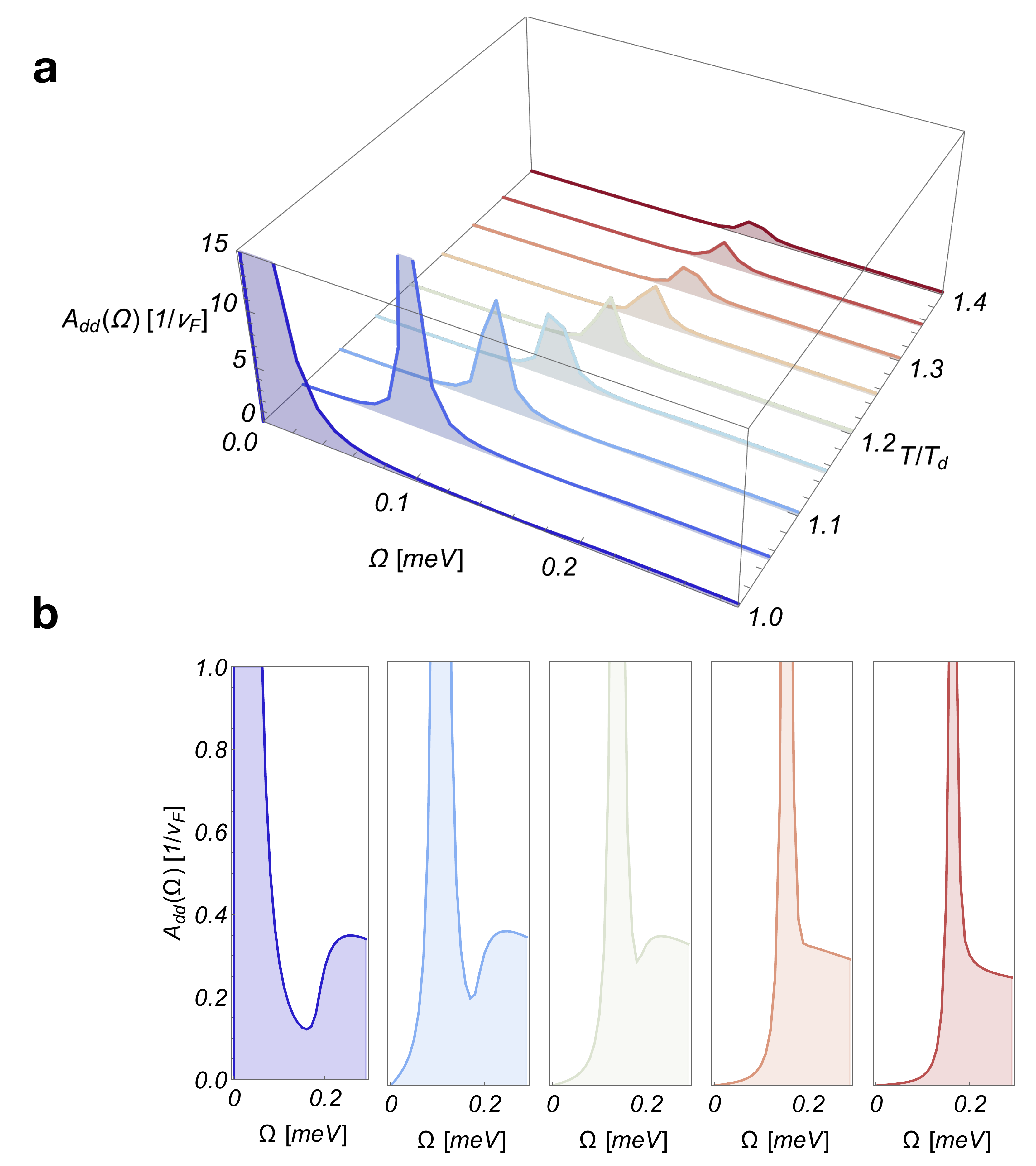}
    \caption{Evolution of the collective mode spectral function with temperature. 
    We fix the tunneling strength $\gamma = .2$ meV and substrate gap to be $\Delta_s = 1.0$ meV, and hold the cutoff $\Lambda = 30$ meV and BCS constant $\lambda^{-1}_d = 4.58658$, corresponding to an intrinsic critical temperature of $T_d^{(0)} = .344$ meV.
    The finite $\gamma$ leads to a reduced critical temperature of $T_d = .282$ meV, giving ratio $T_d^{(0)} \sim 1.2 T_d$.
    \textbf{a.} The collective mode spectral function for temperatures between $T_d$ and $1.4\, T_d$, from which we see that the spectral peak of the mode sharpens and progressively softens as the temperature is lowered. At $T = T_d$, the mode ultimately softens to zero frequency. \textbf{b.} The spectral function for $T/T_d = 1, 1.1, 1.2, 1.3, 1.4$. We see that the mode clearly separates from the quasiparticle continuum below $T \lesssim T_d^{(0)}$, and softens completely at $T_d$.}
    \label{fig:bs-temp}
\end{figure}

We present the spectral function $\mathcal{A}_{dd}(\Omega,\mathbf{q}=0)$ in Fig.~\ref{fig:bs-temp} for different temperatures $T\geq T_d$ at fixed $\gamma, \Delta_s,T_d^{(0)}$.
At high temperatures, we see no clear distinction between the collective mode and the bottom of the quasiparticle continuum.
Lowering the temperature reduces thermal broadening and pulls the mode out of the continuum, yielding a sharp collective mode which resides within the proximity-induced minigap.
This occurs once the temperature $T \sim T_d^{(0)}$, the characteristic temperature scale of the intrinsic pairing interactions.
As we decrease the temperature further the mode continues to soften.
Remarkably, at $T = T_d<T_d^{(0)}$ the mode softens completely, and we see the $d$-wave bound-state itself condenses. 
As we now demonstrate, this signals the onset of a second phase transition into a state with finite $d$-wave order.


\begin{figure}
    \centering
    \includegraphics[width=\linewidth]{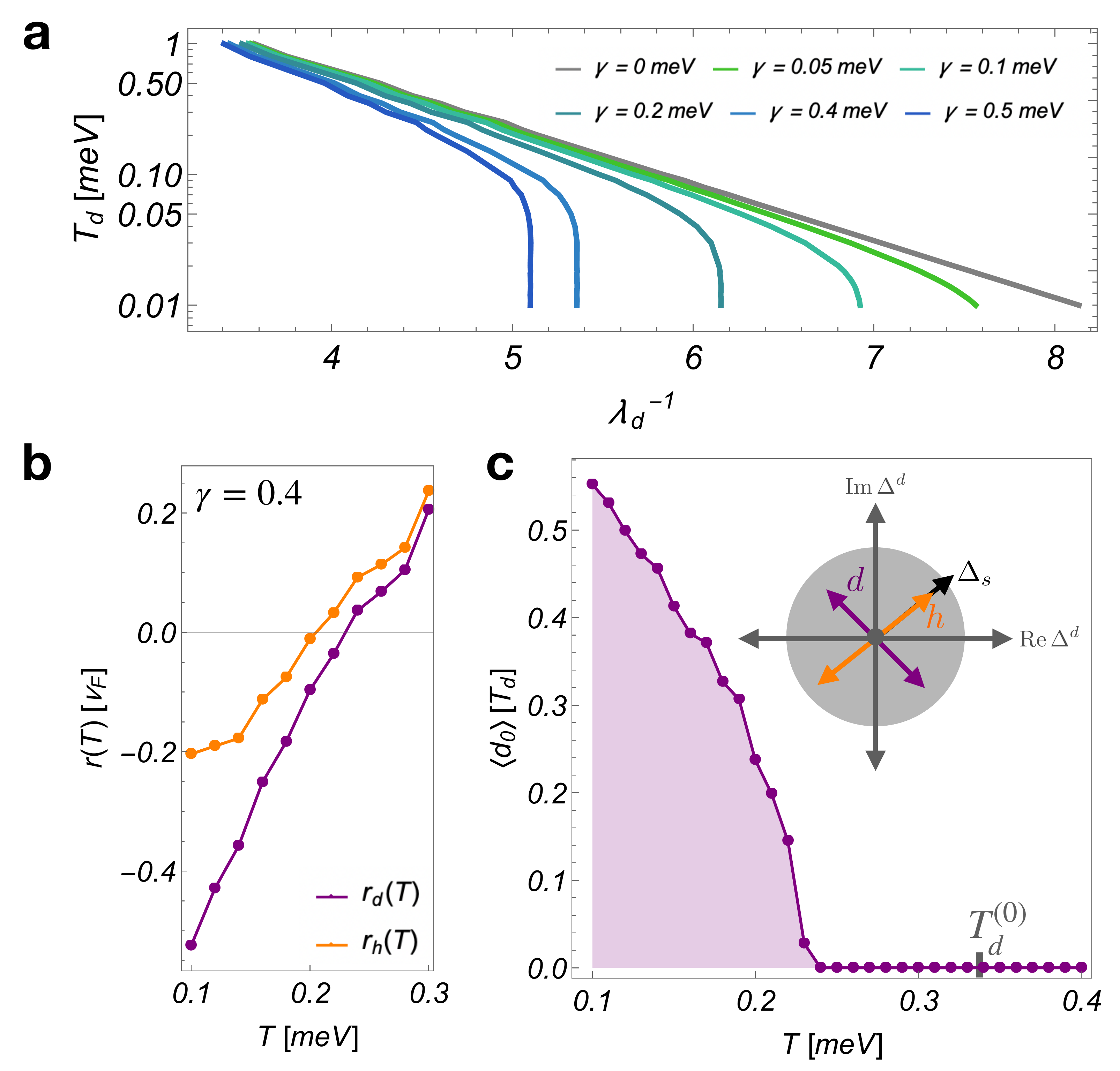}
    \caption{\textbf{a.} Transition temperature $T_d$ of the sample  as a function of the dimensionless $d$-wave coupling constant $\lambda_d = \nu_F g_d$, plotted for different values of the tunneling rate $\gamma$. Notice that for any fixed $\lambda_d$, $T_d$ is suppressed as the tunneling rate is increased. \textbf{b.} Temperature dependence of the quadratic coefficients in the Ginzburg-Landau expansion for $h$ and $d$. The change of sign of each coefficient signals condensation in that channel, and one sees that the out-of-phase $d$ mode condenses first. \textbf{c.} Amplitude of the sample order parameter $\Delta^d = i\langle d \rangle$ near $T_d$. Inset: illustration of the relative Higgs and Bardasis-Schreiffer modes, and their phase relative to the $s$-wave substrate order parameter}
    \label{fig:tc}
\end{figure}

In fact, this split transition behavior is a generic feature of systems with strongly competing superconducting orders, and in this case it signals the onset of time-reversal symmetry breaking \cite{Sigrist2000,hirschfeld-s+id,mixedsymm}.
This is understood by noting that the collective mode is in the out-of-phase channel, and therefore it is odd under time-reversal symmetry. 
Condensing this mode requires spontaneously choosing the relative phase to be either $+\pi/2$ or $-\pi/2$, entering into either an $s+id$ or $s-id$ state~\cite{hirschfeld-s+id,mixedsymm}.

This intuition is confirmed by explicitly solving the Ginzburg-Landau mean-field equation for the $d$-mode as we pass through the temperature $T_d$.
Expanding the gap equation Eq.~\ref{eqn:d-wave-gap} for small $\Delta_d$ we obtain an equation for the static, homogeneous component of $d_{q=0} \equiv d$ (see Supplemental Material) of 
\begin{equation}
    \left(r_d + u_d d^2 \right) d = 0 . 
\end{equation}
The coefficients $r_d  \sim T - T_d $ and $u_d\sim 1/T^2$~\cite{agd}, as well as the quadratic Ginzburg-Landau coefficient for the amplitude mode $r_h$ are calculated microscopically in the Supplemental Material.
In Fig.~\ref{fig:tc}(b), we plot the coefficient $r_d$ alongside $r_h$ and see explicitly that $r_d$ changes sign first at $T = T_d$, confirm our inference based on the collective mode spectrum.
Below $T_d$ the order parameter $d$ acquires a non-zero value shown explicitly in Fig.~\ref{fig:tc}(c). 
Note this transition does not spontaneously break $U(1)$ symmetry, which has already been broken by the substrate order parameter, but it does break the remnant $\mathbb{Z}_2$ symmetry, under which $id \to - id $.

The breaking of time-reversal symmetry in such an ``$s$-$d$" heterostructure has already been predicted, for instance at the twin-grain boundaries in cuprate systems~\cite{sigrist-s-d,sigrist-trsb-3,sigrist-trsb-2}, at the interface of ``$s$-$d$" superconductors~\cite{Huck.1997,sigrist-trsb-1,Lin.2012}, and between twisted cuprate layers~\cite{twisted-bscco,twisted-bscco-2,Zhao.2021}.
Our calculation seems to indicate that the breaking of time-reversal symmetry in these systems is in some sense the ``final fate" of what we have discussed here, and in particular the breaking of time-reversal symmetry in these systems ought to be heralded by a softening collective mode, as we have shown. 
It is also worth commenting that, just as we have shown the Bardasis-Schrieffer collective mode emerges in the normal state of the heterostructure, we may also expect a new collective mode to emerge once time-reversal symmetry is broken below $T_d$~\cite{poniatowski2021spectroscopic,214-clapping,Balatsky.2000,Marciani.2013}, thereby making connection to previous proposals for collective-mode spectroscopy.

\begin{figure}
    \centering
    \includegraphics[width=\linewidth]{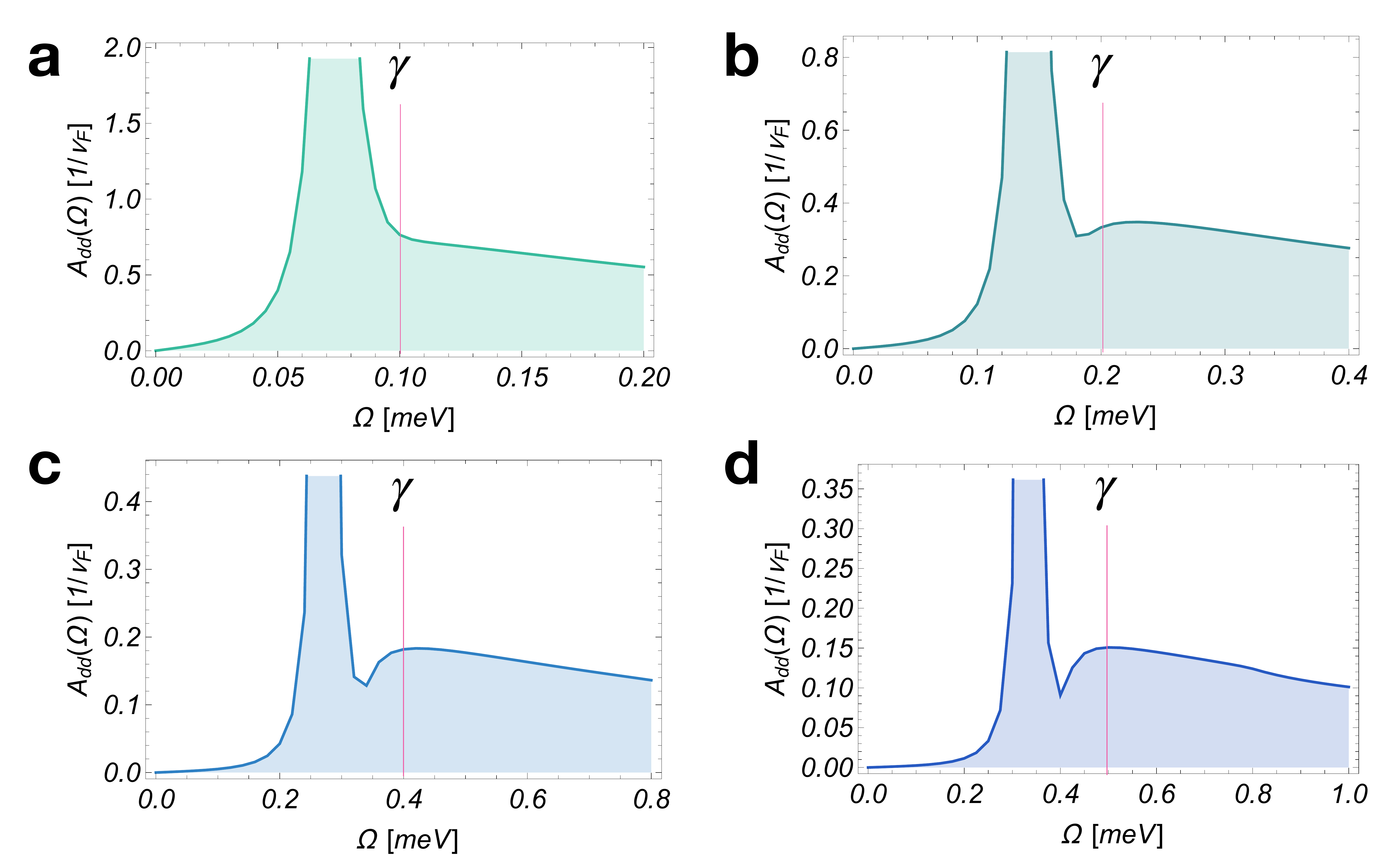}
    \caption{Spectral function of proximity-induced Bardasis-Schrieffer collective mode for different strengths of the tunneling-induced minigap. 
    We fix the substrate gap to be $\Delta_s = 1.0$ meV, and hold the cutoff $\Lambda = 30$ meV and BCS constant $\lambda^{-1}_d = 4.58658$, corresponding to an intrinsic critical temperature of $T_d^{(0)} = .344$ meV.
    We also fix the temperature $T = T_d^{(0)}$.
    We then study the spectral functions while we vary the size of the tunneling rate $\gamma$ from $.1$meV in (a) through $.4$meV in (d) in increments of $.1$ meV. 
    The plots are made over a frequency range from $0$ to $2\gamma$, and at $\Omega = \gamma$ we place a line as a guide to the eye, which indicates where the two-particle continuum formally begins. }
    \label{fig:bs-gamma}
\end{figure}

Before concluding we study the dependence of the collective mode found in Fig.~\ref{fig:bs-temp} upon the minigap $\gamma$.
The minigap $\gamma$ is crucial for the existence of the collective mode, as otherwise it would be overdamped by the quasiparticle continuum.
In Fig.~\ref{fig:bs-gamma} we illustrate this explicitly by plotting the spectral function $\mathcal{A}_{dd}(\Omega,\mathbf{q}=0)$ for a variety of tunneling strengths $\gamma$ at fixed temperature $T = T_d^{(0)}$ and substrate gap $\Delta_s$.
For small minigap $\gamma$, seen in Fig.~\ref{fig:bs-gamma}(a), we see the collective mode is indistinguishable from the quasiparticle continuum\footnote{In particular, in the mean-field limit the two-particle spectral function is completely fixed by the single-particle density-of-states, and this exhibits a weak singularity at the gap edge which may be confused with a collective mode if the resonance lies to close to the continuum.}.
Proceeding to larger values of $\gamma$, e.g. in Fig.~\ref{fig:bs-gamma}(c),(d), we see a well separated mode appears within the minigap.
Practically, this points to the necessity of establishing good electrical contact between the sample and substrate, so that the tunneling scale, and therefore minigap $\gamma$, is as large as possible.

We also briefly comment here on the $\Delta_s$ dependence of the mode, which is studied in the Supplementary Material.
In particular, we see that at finite $\gamma$ and small $\Delta_s$ the substrate acts as an incoherent reservoir. 
In this case we recover the known behavior for overdamped fluctuations of the sample's superconducting order~\cite{larkin-book}, with the $d$-wave pairs decaying with a characteristic lifetime $\tau \sim 1/(T - T_d)$, thereby also allowing for the study of critical superconducting fluctuations~\cite{Scalapino.1970,Kadin.1982,Goldman.2006}.
Nevertheless, for the purposes outlined in this work, we see that it is beneficial for the substrate to have as large a gap as possible. 
We now briefly discuss various experimental signatures of this mode.


The first is electron tunneling spectroscopy~\cite{Wolf}. 
In inelastic tunneling spectroscopy, bosonic excitations such as phonons~\cite{Vitali.2004} and magnons~\cite{Balashov.2006} are routinely observed by studying characteristic $I-V$ curves.
In particular, these bosonic excitations may appear as a sharp feature in $d^2I/dV^2$, which signals the opening of a new inelastic scattering channel for electrons at that bias energy. 
In this context, we may imagine it is also possible for an electron to emit a collective mode in the process of tunneling in to the sample, and therefore we should also expect a similar kink feature to appear in the $I-V$ curve once the energy passes the collective mode threshold. 
While this is still a relatively difficult measurement to perform, there is some precedent for using this technique to study collective modes of unconventional superconductors~\cite{Hlobil.2017,Jandke.2016}. 
Since, as we have seen, the collective mode we identify here can have strong temperature dependence, this may help to identify such a feature since it in principle will soften considerable as the temperature is lowered. 

In a similar vein, it may also be possible to identify this collective mode using ARPES, in which case the mode will again manifest as an inelastic contribution to the electronic self-energy~\cite{Norman.1999,Valla.1999,Hoffman.2009}. 
In this context, ARPES has the additional benefit of potentially observing the momentum dependence of the coupling, which could help identify the symmetry channel of the collective mode, and therefore identify the symmetry channel of the underlying pairing interaction. 

Lastly, we also expect Raman spectroscopy to be sensitive to the collective mode.
This is not surprising since it has long been known that the Bardasis-Schrieffer mode, when it exists, is a Raman active mode~\cite{bs-raman,Devereaux.2007,Scalapino.2009}.
Like ARPES, Raman spectroscopy also has the potential to probe the selection rules of the collective mode in addition to the frequency of the mode. 

Finally, we discuss promising materials for the realization of this proposal.
In our model, we considered a $d$-wave system but this is not crucial; much of what we assumed only relied on the sample order parameter being orthogonal to the $s$-wave substrate order. 
However, we do want the intrinsic critical temperature $T_d^{(0)}$ to be low compared to the bulk transition temperature of the substrate, $T_s$. 

There are a number of interesting van der Waals compounds~\cite{Zhou.2016,Wakatsuki.2016}, such as MoS$_2$~\cite{Lu.2015,Saito.2016}, NbSe$_2$~\cite{Xi.2016}, WS$_2$~\cite{Lu.2018}, and WTe$_2$~\cite{wte2,wte2-1,fu-wte2} which exhibit possibly unconventional superconductivity and can be exfoliated into thin layers.
In addition, moir\'e bilayer and trilayer graphene quite likely exhibit unconventional superconductivity at a low temperature scale and are also easily exfoliated~\cite{bilayer-1,trilayer-1,bilayer-stm,trilayer-stm}. 

It may also be possible to study this physics using a severely overdoped cuprate, provided the transition temperature can be depressed below that of a realistic $s$-wave system. 
This has the benefit of having an established gap-symmetry and therefore may offer a useful test-case. 
In addition, recent efforts have established that certain cuprates may also be prepared in thin layers, or even single copper-oxide layers~\cite{Yu.2019}.
In this context, our proposal has some technical overlap with recent proposals for time-reversal symmetry breaking chiral superconductivity in systems of twisted cuprate monolayers~\cite{twisted-bscco,twisted-bscco-2,Zhao.2021}.

The main limitation is the requirement that the $s$-wave substrate superconductor have a higher critical temperature than the sample.
To maximize the substrate transition temperature, the most likely candidate substrates are then Nb, NbN, or NbTiN, with  $T_s\sim 7-15$ K.
It may also be possible to use a fullerene such as Rb$_3$C$_{60}$, which would allow one to access higher temperatures $T_s \sim 30$K, at the expense of likely introducing other complications~\cite{Tanigaki.1991}.

It would be interesting to try and apply our results to sample superconductors which already feature intrinsic collective modes but which are overdamped, e.g. due to nodal quasiparticles. 
By opening a proximity-induced gap, one may attempt to, e.g. stabilize the Higgs collective mode, which is usually located at the gap edge and subject to quasiparticle damping~\cite{higgs-cdf}.
In this way, much like a charge-density wave order parameter can separate the Higgs mode from the continuum and enable its coherent oscillation \cite{higgs-arcmp,varma-cdw,higgs-cdw}, it might also be possible to use the small proximity-induced gap to separate the continuum from the Higgs mode and enable its widespread detection. 

In conclusion, we have considered a simple model of an unconventional superconducting sample that is proximitized by a dominant $s$-wave superconducting substrate and seen that this can lead to a sharp collective mode which captures the intrinsic pairing interaction in the sample.
This then greatly expands the platforms for studying unconventional superconductivity through their collective modes and increases the number of experimentally accessible probes for these difficult to characterize states.

\section{Acknowledgements}

We acknowledge useful discussions with Eugene Demler, Stuart S.P. Parkin, Manfred Sigrist, Matteo Mitrano, Ken Burch, Zachary Raines, Andrew Allocca, Niels Schr\"oter, and Mostafa Marzouk. 
N.R.P. and J.B.C. acknowledge the hospitality of the ETH Z\"urich Institute for Theoretical Physics and the Max Planck Institute for the Structure and Dynamics of Matter (MPSD, Hamburg), where part of this work was completed. 
This work was primarily supported by the Quantum Science Center (QSC), a National Quantum Information Science Research Center of the U.S. Department of Energy (DOE). 
J.B.C. is an HQI Prize Postdoctoral Fellow and gratefully acknowledges support from the Harvard Quantum Initiative.
 N.R.P. is supported by the Army Research Office through an NDSEG fellowship. 
  A.Y. is partly supported by 
the Gordon and Betty Moore Foundation through Grant GBMF 9468 and by the National Science Foundation under Grant No. DMR-1708688. P.N. is grateful for the hospitality of the Max Planck Institute for the Structure and Dynamics of Matter where part of this work was completed supported by a Max Planck Sabbatical Award and a Bessel Research Award of the Alexander von Humboldt Foundation.
P.N. is a Moore Inventor Fellow and gratefully acknowledges support through Grant GBMF8048 from the Gordon and Betty Moore Foundation.

\bibliography{references}

\clearpage
\onecolumngrid
\appendix*


\section{Substrate Self-Energy}
\label{sec:substrate}

Here we derive the self-energy for the sample electrons and in particular recover the proximity effect Hamiltonian in the regime dominated purely by Andreev processes. 

For a detailed treatment see, for instance, Ref.~\onlinecite{patrick-lee-sign,majorana-local,Stanescu.2010}.
We begin by employing the Matsubara framework and model the tunneling interaction via 
\begin{equation}
    \mathcal{S}_{\rm int} = -\mathfrak{t}\int d^2 r \int d\tau \left[ \overline{\psi}(x) \tau_3 \Psi(x,z = 0) +  \overline{\Psi}(x,z=0) \tau_3 \psi(x) \right].
\end{equation}
Here $\mathfrak{t}$ is an effective local, spin and momentum independent tunneling matrix element, and $\psi(x)$ is used to describe the electrons in the thin-layer sample while $\Psi(x,z)$ describes the electrons in the substrate with depth $z \leq 0$ (the interface is taken to be at $z =0$).

We can formally integrate out the substrate electrons assuming a Gaussian approximation, which is well-justified if the phase fluctuations are frozen out. 
We then generate an effective action for the sample electrons of 
\begin{equation}
    \mathcal{S}_{\rm eff} = - \log \langle e^{-S_{\rm int}} \rangle
\end{equation}
with the expectation value evaluated using the substrate Green's function.
We find the formal result 
\begin{equation}
    \mathcal{S}_{\rm eff} = -\mathfrak{t}^2 \int d^3 x d^3 x' \overline{\psi}(x')\tau_3 \langle \Psi(x',z=0)\overline{\Psi}(x,z=0) \rangle \tau_3 \psi(x) ,
\end{equation}
or in terms of the substrate Green's function 
\begin{equation}
    \mathcal{S}_{\rm eff} = \mathfrak{t}^2 \int d^3 x d^3 x' \overline{\psi}(x')\tau_3 \mathbb{\hat{G}}_{\rm sub}(x',z=0;x,z=0) \tau_3 \psi(x) .
\end{equation}
We use the well-known ``local approximation" which evaluates the substrate Green's function locally in space via 
\begin{equation}
    \mathbb{\hat{G}}_{\rm sub}(x',z=0;x,z=0) \sim \mathbb{\hat{G}}_{\rm sub}(\tau',\mathbf{r};\tau,\mathbf{r})\delta^2(\mathbf{r}'-\mathbf{r}).
\end{equation}
This is then related to the local density-of-states in the substrate in the frequency domain as 
\begin{equation}
\mathbb{\hat{G}}_{\rm sub}(i\varepsilon_{m};\mathbf{r},\mathbf{r}) = \int_{\bf p} \left( i\varepsilon_m - \xi_{\bf p}\tau_3 - \Delta_s \tau_1 \right)^{-1} = -\pi \nu_s \frac{i\varepsilon_m + \Delta_s \tau_1}{\sqrt{\varepsilon_m^2 + \Delta_s^2}}.
\end{equation}
In the effective action for the sample, this means that we find a contribution from the substrate of 
\begin{equation}
    \mathcal{S}_{\rm eff} = \sum_{p} \overline{\psi}_p \bm{\Sigma}_s(p)\psi_p 
\end{equation}
with self-energy 
\begin{equation}
    \bm{\Sigma}_s(p) = \mathfrak{t}^2 \tau_3 \mathbb{\hat{G}}_{\rm sub}(i\varepsilon;\mathbf{r},\mathbf{r}) \tau_3 = -\pi \nu_s \mathfrak{t}^2  \tau_3 \frac{i\varepsilon_m + \Delta_s \tau_1}{\sqrt{\varepsilon_m^2 + \Delta_s^2}}\tau_3 \equiv   -\frac{\gamma}{2} \frac{i\varepsilon_m - \Delta_s \tau_1}{\sqrt{\varepsilon_m^2 + \Delta_s^2}}. 
\end{equation}
This defines the tunneling scale as 
\begin{equation}
    \gamma = 2\pi \nu_s |\mathfrak{t}|^2 .
\end{equation}
It is common to characterize the substrate-effect in terms of the quasiparticle and gap renormalizations via 
\begin{subequations}
\begin{align}
& Z(i\varepsilon_m) = 1 + \frac{\gamma}{2}\frac{1}{\sqrt{\varepsilon_m^2 + \Delta_s^2}}\\
& \Phi(i\varepsilon_m) =  \frac{\gamma}{2}\frac{\Delta_s}{\sqrt{\varepsilon_m^2 + \Delta_s^2}},
\end{align}
\end{subequations}
such that the electronic Green's function in the normal state of the sample is 
\begin{equation}
    \mathbb{\hat{G}}^{-1}_{\rm sample}(i\varepsilon_m,\mathbf{p}) = Z(i\varepsilon_m) i\varepsilon_m - \xi_{\bf p}\tau_3 - \Phi(i\varepsilon_m)\tau_1 . 
\end{equation}
We can also analytically continue this result to get the retarded self-energy via 
\begin{subequations}
\begin{align}
& Z_R(\varepsilon) = 1 + \frac{\gamma}{2}\frac{1}{\sqrt{ \Delta_s^2 - (\varepsilon+ i 0^+)^2 }}\\
& \Phi_R(\varepsilon) =  \frac{\gamma}{2}\frac{\Delta_s}{\sqrt{ \Delta_s^2 - (\varepsilon+ i 0^+)^2 }}.
\end{align}
\end{subequations}
We note that in the limit of $\Delta_s \to \infty$ the quasiparticle renormalization becomes trivial and the anomalous term becomes the minigap, such that 
\begin{subequations}
\begin{align}
& Z_R(\varepsilon) \to 1 \\
& \Phi_R(\varepsilon) =  \to \frac{\gamma}{2}\tau_1.
\end{align}
\end{subequations}
This motivates a BdG Hamiltonian in this regime, dominated by the Andreev reflection back in to the sample, with proximity induced gap term. 

\section{Gap Equation}
In this section, we discuss the mean-field properties of our model in the Matsubara imaginary time formalism. In particular, we solve the gap equation determining the transition temperature of the sample, $T_d$, and show that the resulting state of the coupled sample-substrate system spontaneously breaks time-reversal symmetry, with the sample order parameter condensing $\pi/2$ out-of-phase with the substrate order parameter, forming an $s+id$ state. 

Taking into account the self-energy contribution from the coupling to the substrate, the Matsubara action for the sample reads
\begin{equation} \label{eq:matsubara-act}
    S = \frac{1}{g_d}\sum_q \bar{\Delta}_q^d \Delta_q^d - \tr \log \mathbb{G}^{-1} \, ,
\end{equation}
where $\Delta_q^d$ is the $d$-wave order parameter in the sample and the inverse Gor'kov Green function is 
\begin{equation}
    \hat{\mathbb{G}}^{-1}(p,q) = Z_n i\w_n - \xi_{\p} \tau_3 - \big( \Phi_n + \Delta_q^d \chi_{\v{p}}^d \big) \tau^\dag + \text{h.c.} 
\end{equation}
where $\w_n = 2\pi(n+\frac12)T$ is a fermionic Matsubara frequency and $\tau_i$ are the Pauli matrices in Nambu space, with $\tau = \frac12 (\tau_1 - i\tau_2)$. The quasiparticle renormalization and anomalous self energy due to the substrate are
\begin{align}
    Z_n &= 1 + \frac{\gamma}{2} \frac{1}{\sqrt{\w_n^2 + \Delta_s^2}} \\
    \Phi_n &= -\frac{\gamma}{2} \frac{\Delta_s}{\sqrt{\w_n^2 + \Delta_s^2}} \, .
\end{align}
The BCS gap equation for the homogeneous order parameter $\Delta^d \equiv \Delta_{q=0}^d$ is given by the saddle point of this action,
\begin{equation}
    \Delta^d = - g_d T \sum_p \chi_{\p}^d \, \tr \,\hat{\mathbb{G}}(p,0) \tau = g_d T \sum_p \chi_{\p}^d \, \frac{\Delta^d \chi_{\p}^d + \Phi_n}{Z_n^2 \w_n^2 + \Phi_n^2 + |\Delta^d|^2 \left(\chi_{\p}^d\right)^2} \, .
\end{equation}
The critical temperature $T_d$ of the sample can be determined by solving the gap equation in the limit $\Delta_d \rightarrow 0$. In this limit, the gap equation reduces to 
\begin{equation} \label{eq:matsubara-tc}
    \lambda^{-1}_d = 2\pi T_d \sum_{\w_n < \Lambda} \frac{1}{\sqrt{Z_n^2 \w_n^2 + \Phi_n^2}} = 2\pi T_d \sum_{\w_n < \Lambda} \left[ \left( 1 + \frac{\gamma}{\sqrt{\w_n^2 + \Delta_s^2}} \right) \w_n^2 + \frac{\gamma^2}{4} \right]^{-1/2} \, ,
\end{equation}
where $\lambda_d = \nu_F g_d$ is the dimensionless $d$-wave coupling constant (with $\nu_F$ the density of states at the Fermi level) and we have written $\sum_p = \nu_F \sum_{\w_n}  \int \df \xi \int \frac{\df \theta_{\p}}{2\pi}$ and performed the integrals over $\theta_{\p}$ and $\xi$. The frequency cutoff $\Lambda$ can be expressed in terms of a dimensionless cutoff $N$ on the Matsubara index as $\Lambda = 2\pi N T_d$. 

We can approach the problem from a complementary perspective by expanding the action (\ref{eq:matsubara-act}) in powers of $\Delta^d$, which furnishes an effective Ginzburg-Landau theory, valid near $T_d$. As discussed in the main text, it is useful to decompose $\Delta^d$ into its components in-phase and out-of-phase with the substrate order parameter, writing $\Delta^d = h+id$. In this section, we will be concerned with only the static, homogeneous order parameter at the level of mean field theory, and thus neglect the frequency and momentum dependence of the fields $h$ and $d$.

To organize the expansion, we write the Gor'kov Green function as $\hat{\mathbb{G}}^{-1} = \hat{\mathbb{G}}^{-1}_0 + \hat{\Lambda}^h h + \hat{\Lambda}^d d$, with $\hat{\Lambda}^h = -\chi_{\p}^d \tau_1$ and $\hat{\Lambda}^d = \chi_{\p}^d \tau_2$. Expanding (\ref{eq:matsubara-act}) to fourth order in $h$ and $d$, we find
\begin{equation}
    S = r_h \, h^2 + r_d d^2 + u_h h^4 + u_d d^4 + u' d^2 h^2 \, .
\end{equation}
The superconducting transition occurs when $r_d$ or $r_h$ changes sign, signalling an instability in the in-phase (nematic) or out-of-phase (time reversal symmetry breaking) channel. Both of these functions are related to the (inverse) fluctuation propagator $L_{dd}^{-1}(q=0)$ discussed in the main text, and are explicitly given by 
\begin{align}
    r_h(T) &=g_d^{-1} + \frac{T}{2}\sum_p \left(\chi_{\p}^d\right)^2 \, \tr \left(\hat{\mathbb{G}}_0 \hat{\Lambda}^h\right)^2 = g_d^{-1} - 2\pi \nu  T\sum_{\w_n < \Lambda} \frac{1}{\sqrt{Z_n^2 \w_n^2 + \Phi_n^2}} + \pi \nu T \sum_{\w_n < \Lambda} \frac{\Phi_n^2}{\left[Z_n^2 \w_n^2 + \Phi_n^2 \right]^{3/2}}\\
    r_d(T) &= g_d^{-1} + \frac{T}{2}\sum_p \left(\chi_{\p}^d\right)^2 \, \tr \left(\hat{\mathbb{G}}_0 \hat{\Lambda}^d\right)^2 = g_d^{-1} - 2\pi \nu  T\sum_{\w_n < \Lambda} \frac{1}{\sqrt{Z_n^2 \w_n^2 + \Phi_n^2}} \, .
\end{align}
Clearly, the zeroes of $r_d(T)$ coincide with the solutions to (\ref{eq:matsubara-tc}) which determine $T_d$. Moreover, one finds numerically that the second term in $r_h$ above is always positive, so that $r_d$ always changes sign first (i.e., before $r_h$) as the temperature is lowered. This implies that the sample order parameter condenses out-of-phase with the substrate order parameter, $\Delta^d \sim id$ which implies the system spontanously breaks time reversal symmetry at $T_d$.

To stabilize the expansion in $d$, we must calculate the quartic coefficient $u_d$, which is given by 
\begin{equation}
    u_d = \frac{3\nu}{8}\, 2\pi T \sum_{\w_n < \Lambda} \frac{1}{\left[ Z_n^2 \w_n^2 + \Phi_n^2\right]^{3/2}} \, .
\end{equation}
We may then solve the saddle point equation for $d$, as discussed in the main text, which allows us to determine the equilibrium value of the sample order parameter near $T_d$,
\begin{equation}
    \langle \Delta^d \rangle = i\langle d \rangle = i \sqrt{\frac{-r_d(T)}{2u_d}} \, .
\end{equation}

\section{\label{sec:keldysh}Collective Mode Propagator}
Here we derive the collective mode propagators in the Random Phase Approximation using the Keldysh technique.

We follow Kamenev~\cite{kamenev} and introduce fermion fields on the $\pm$ time contours.
The BCS action can be written in terms of Nambu-Gorkov space for each time-contour as 
\begin{multline}
      S = \int dt \sum_{\bf p}\overline{\psi}_{\bf p}\sigma_3 \left[ i\partial_t - \xi_{\bf p}\tau_3 \right] \psi_{\bf p}\\
      +g_d \sum_{\bf q}\int_{\bf p,p'} \chi_{\bf p}^d \chi_{\bf p'}^d \int dt \left[ \overline{\psi}_{{\bf p}+\frac12\bf q + } \tau^\dagger {\psi}_{{\bf p}-\frac12\bf q + }\overline{\psi}_{{\bf p}'-\frac12\bf q + } \tau {\psi}_{{\bf p}'+\frac12\bf q + } - \overline{\psi}_{{\bf p}+\frac12\bf q - } \tau^\dagger {\psi}_{{\bf p}-\frac12\bf q - }\overline{\psi}_{{\bf p}'-\frac12\bf q - } \tau {\psi}_{{\bf p}'+\frac12\bf q - } \right]  
\end{multline}
Here $\tau$ are the Nambu-Gor'kov matrices and $\sigma$ are the Keldysh matrices.
We perform the Larkin-Ovchinikov rotation 
\begin{equation}
    \begin{pmatrix}
    \psi_{\bf p + } \\
    \psi_{\bf p -} \\
    \end{pmatrix} = \frac{1}{\sqrt{2}}(\sigma_1 + \sigma_3 )    \begin{pmatrix}
    \psi_{{\bf p} S } \\
    \psi_{{\bf p} A } \\
    \end{pmatrix},\quad \left(\overline{\psi}_{\bf p +},\overline{\psi}_{\bf p -} \right) = \left(\overline{\psi}_{{\bf p} S},\overline{\psi}_{{\bf p} A} \right)\frac{1}{\sqrt{2}}(\sigma_1 + \sigma_3)\sigma_3.
\end{equation}
we henceforth use $\psi$ to indicate the rotated spinor. 

We find action 
\begin{multline}
      S = \int dt \sum_{\bf p}\overline{\psi}_{\bf p}\left[ i\partial_t - \xi_{\bf p}\tau_3 \right] \psi_{\bf p}\\
      +g_d \sum_{\bf q}\int_{\bf p,p'} \chi_{\bf p}^d \chi_{\bf p'}^d \int dt \left[ \overline{\psi}_{{\bf p}+\frac12\bf q } \frac{\sigma_1}{2} \tau^\dagger {\psi}_{{\bf p}-\frac12\bf q}\overline{\psi}_{{\bf p}'-\frac12\bf q } \tau {\psi}_{{\bf p}'+\frac12\bf q } + \overline{\psi}_{{\bf p}+\frac12\bf q } \tau^\dagger {\psi}_{{\bf p}-\frac12\bf q}\overline{\psi}_{{\bf p}'-\frac12\bf q } \frac{\sigma_1}{2} \tau {\psi}_{{\bf p}'+\frac12\bf q } \right]  .
\end{multline}

We perform a Hubbard-Stratonovich decoupling of the interaction in the Cooper channel. 
We introduce fields $\Delta^{cl}_{\bf q}(t), \Delta^{q}_{\bf q}(t)$ and their conjugates and write the action as 
\begin{multline}
    S = \int dt \sum_{\bf p}\overline{\psi}_{\bf p}\left[ i\partial_t - \xi_{\bf p}\tau_3 \right] \psi_{\bf p} \\
    -\sum_{\bf q}\int_{\bf p}\int dt\left[ \overline{\psi}_{{\bf p}-\frac12{\bf q}}\chi_{\bf p}^d \overline{\Delta}^{cl}_{\bf q}\tau \psi_{{\bf p}+\frac12{\bf q}} + \overline{\psi}_{{\bf p}+\frac12{\bf q}}\chi^d_{\bf p}{\Delta}^{cl}_{\bf q}\tau^\dagger \psi_{{\bf p}-\frac12{\bf q}} \right]\\
    +\int dt \sum_{\bf q}\int_{\bf p} \overline{\Delta}_{\bf q}^{q}\left[ \frac{-1}{g_d}\Delta^{cl}_{\bf q} - \int_{\bf p}\chi^d_{\bf p}\overline{\psi}_{{\bf p}-\frac12{\bf q}} \frac{\sigma_1}{2}\tau \psi_{{\bf p}+\frac12{\bf q}} +\right]+{\Delta}_{\bf q}^{q}\left[ \frac{-1}{g_d}\overline{\Delta}^{cl}_{\bf q} - \int_{\bf p}\chi^d_{\bf p}\overline{\psi}_{{\bf p}+\frac12{\bf q}} \frac{\sigma_1}{2}\tau^\dagger \psi_{{\bf p}-\frac12{\bf q}} \right].
\end{multline}
Here we see the field $\Delta^q$ acts a Lagrange multiplier for the classical field.

We simplify to get 
\begin{equation}
    S = \int dt \int_{\bf p}\sum_{\bf q} \overline{\psi}_{{\bf p}+\frac12{\bf q}}\left[ \delta_{{\bf q},0}(i\partial_t - \xi_{\bf p}\tau_3) - \tau^\dagger\chi_{\bf p}^d(\frac{\sigma_1}{2} \Delta^q_{\bf q} + \Delta^{cl}_{\bf q} ) -\tau\chi_{\bf p}^d(\frac{\sigma_1}{2} \overline{\Delta}^q_{-\bf q} + \overline{\Delta}^{cl}_{-\bf q} ) \right] \psi_{{\bf p}-\frac12\bf q } - \sum_{\bf q} \int dt \frac{1}{g_d}\left( \overline{\Delta}^{q}_{\bf q}\Delta^{cl}_{\bf q} +  \overline{\Delta}^{cl}_{\bf q}\Delta^{q}_{\bf q}\right).
\end{equation}
We can write this compactly by introducing the Keldysh kernel
\begin{equation}
   \mathbb{\check{G}}^{-1}_{p,q} = \delta_{q,0}\left( \varepsilon - \xi_{\bf p}\tau_3\right) - \chi^d_{\bf p}\left( \tau^\dagger \check{\Delta}_{q} + \tau \check{\overline{\Delta}}_{-q}\right)
\end{equation}
with the pair scattering vertices 
\begin{equation}
   \check{\Delta}_{q} = \frac{\sigma_1}{2} \Delta^q_{q} + \Delta^{cl}_q
\end{equation}
where we have introduced the fermionic and bosonic four-momenta $p=(\varepsilon,{\bf p}), q = (\omega,{\bf q})$.

We can now include the effect of the substrate via the retarded self-energy computed above.
We have 
\begin{equation}
    \hat{\bm\Sigma}^R(\varepsilon) = -\frac{\gamma}{2}\left( \frac{\varepsilon +\Delta_s \tau_1}{\sqrt{\Delta_s^2 -(\varepsilon+i0^+)^2 }}\right).
\end{equation}
From this we obtain the advanced self-energy as 
\begin{equation}
    \hat{\bm\Sigma}^A(\varepsilon) = -\frac{\gamma}{2}\left( \frac{\varepsilon +\Delta_s \tau_1}{\sqrt{\Delta_s^2 -(\varepsilon-i0^+)^2 }}\right),
\end{equation}
and Keldysh self-energy via fluctuation-dissipation relation of 
\begin{equation}
    \hat{\bm\Sigma}^K(\varepsilon) = F(\varepsilon)\left(\hat{\bm\Sigma}^R(\varepsilon) - \hat{\bm\Sigma}^A(\varepsilon) \right).
\end{equation}
The function $F(\varepsilon)$ is the Keldysh occupation function and in equilibrium it is fixed to be 
\begin{equation}
    F(\varepsilon) = \tanh(\frac{\beta\varepsilon}{2}).
\end{equation}
This yields the Keldysh kernel of 
\begin{equation}
   \mathbb{\check{G}}^{-1}_{p,q} = \delta_{q,0}\left( \varepsilon - \xi_{\bf p}\tau_3 - \check{\bm \Sigma}(p)\right) - \chi^d_{\bf p}\left( \tau^\dagger \check{\Delta}_{q} + \tau \check{\overline{\Delta}}_{-q}\right), 
\end{equation}
such that the BCS action is 
\begin{equation}
    S = -\frac{1}{g_d} \sum_q \left( \overline{\Delta}^q_{q}\Delta^{cl}_q +  \overline{\Delta}^{cl}_{q}\Delta^{q}_q\right) + \overline{\psi}\cdot \check{\mathbb{G}}\cdot \psi ,
\end{equation}
and the effective action obtained by integrating out the electrons is 
\begin{equation}
    S^{\rm eff} = -\frac{1}{g_d} \sum_q \left( \overline{\Delta}^q_{q}\Delta^{cl}_q+\overline{\Delta}^{cl}_{q}\Delta^{q}_q\right) - i \Tr \log \check{\mathbb{G}}^{-1}[\Delta,\overline{\Delta}]. 
\end{equation}
This functional of the order parameter is then evaluated in a saddle-point expansion. 

\subsection{Saddle-Point}
For the saddle-point we take 
\begin{equation}
    \frac{\delta S^{\rm eff}}{\delta \overline{\Delta}^q_q} = 0 
\end{equation}
and assume $\Delta^{cl}_q$ only has a condensate at zero momentum. 
We find the gap equation 
\[
    -\frac{1}{g_d}\Delta^{cl} - i \int_p \tr  \mathbb{\check{G}}(p)(-\frac{\sigma_1}{2}) \tau \chi^d_{\bf p} = 0. 
\]
This reads 
\begin{equation}
\frac{1}{g_d} \Delta^{cl} = \frac{i}{2}\int_p \tr  \chi^d_{\bf p} \tau \sigma_1 \mathbb{\check{G}}(p).
\end{equation}
The trace over the Keldysh space gives the Keldysh component, such that 
\begin{equation}
\frac{1}{g_d} \Delta^{cl} = \frac{i}{2}\int_p \tr  \chi^d_{\bf p} \tau \mathbb{\hat{G}}^K(p).
\end{equation}
We have 
\begin{equation}
    \mathbb{\hat{G}}^R(p) = \left(Z_R(\epsilon)\epsilon - \xi_{\bf p}\tau_3  - \Phi_R(\varepsilon)\tau_1 - \Delta^{cl}\chi^d_{\bf p}\tau^\dagger - \overline{\Delta}^{cl}\chi^d_{\bf p}\tau \right)^{-1}
\end{equation}
with the wavefunction renormalization and anomalous self-energy 
\begin{equation}
\begin{aligned}
& Z_R = 1 + \frac12\gamma \frac{1}{\sqrt{\Delta_s^2 - (\epsilon+i0^+)^2 }} \\
& \Phi_R(\varepsilon) =  -\frac12\gamma \frac{\Delta_s }{\sqrt{\Delta_s^2 - (\epsilon+i0^+)^2 }} .
\end{aligned}
\end{equation}
The trace over $\tau$ selects the anomalous component, while the integral over the $d$-wave form factor projects out the substrate contribution. 
Crucially, this only holds if the unconventional order is in an orthogonal channel to the $s$-wave substrate order. 
In the limit of small $\Delta^{cl}$, using the fluctuation dissipation representation for $\mathbb{\hat{G}}^K$ we find the linearized gap equation 
\begin{equation}
    \frac{1}{g_d}\Delta^{cl} = \frac{i}{2} \int_p F(\varepsilon) (\chi^d_{\bf p})^2\Delta^{cl}\left[ \frac{1}{ (Z_R(\varepsilon) \varepsilon )^2 -\xi_{\bf p}^2 - (\Phi_R(\varepsilon) )^2  } -\frac{1}{ (Z_A(\varepsilon) \varepsilon )^2 -\xi_{\bf p}^2 - (\Phi_A(\varepsilon) )^2  } \right].  
\end{equation}
We make the quasiclassical approximation and average over the Fermi surface, enabled by the fact that the gap is $s$-wave and respects the symmetry of the Fermi surface.
Introducing density-of-states at the Fermi level $\nu_F$, and pairing constant $\lambda = g_d \nu_F$ we find 
\begin{equation}
    \frac{1}{\lambda} = \int d \varepsilon \frac12 \tanh\frac{\beta\varepsilon}{2} \int d\xi P(\varepsilon,\xi) .  
\end{equation}
with pairing spectral function
\begin{equation}
 P(\varepsilon,\xi) = -\frac{1}{\pi}\Im \left[ \frac{1}{ (Z_R(\varepsilon) \varepsilon )^2 -\xi^2 - (\Phi_R(\varepsilon) )^2  } \right].
\end{equation}
It is easily seen that this is an even function of $\xi$, which is cutoff at $\xi = \Lambda $, and we can see that $\varepsilon \to -\varepsilon$ corresponds to taking the complex conjugate (or alternatively, switches the $R$ and $A$ components), such that this is an odd function of $\varepsilon$. 
We therefore fold the integrations twice, noting $\tanh$ is also odd in $\varepsilon$.
Thus we get 
\begin{equation}
    \frac{1}{\lambda} = 2 \int_0^{\infty} d \varepsilon \tanh\frac{\beta\varepsilon}{2} \int_0^{\Lambda} d\xi P(\varepsilon,\xi) .  
\end{equation}
Note that in the absence of the self-energy we have 
\[
P(\varepsilon,\xi) = \textrm{sgn}(\varepsilon)\delta(\varepsilon^2 - \xi^2),
\]
which gives the equation for $\lambda$ of 
\[
\frac{1}{\lambda} = 2 \int_0^{\Lambda} d\xi \frac{\tanh(\beta\xi/2)}{2\xi} \Rightarrow T_d^{(0)} = \frac{ 2 e^{\gamma_E}}{\pi}\Lambda e^{-1/\lambda},
\]
which is the standard BCS gap equation. 
Here we have introduced the value of $T_d^{(0)}$ which is the intrinsic $d$-wave transition temperature in the absence of the substrate.
In the presence of the substrate, this will be evaluated numerically. 

\begin{figure}
    \centering
    \includegraphics[width=\linewidth]{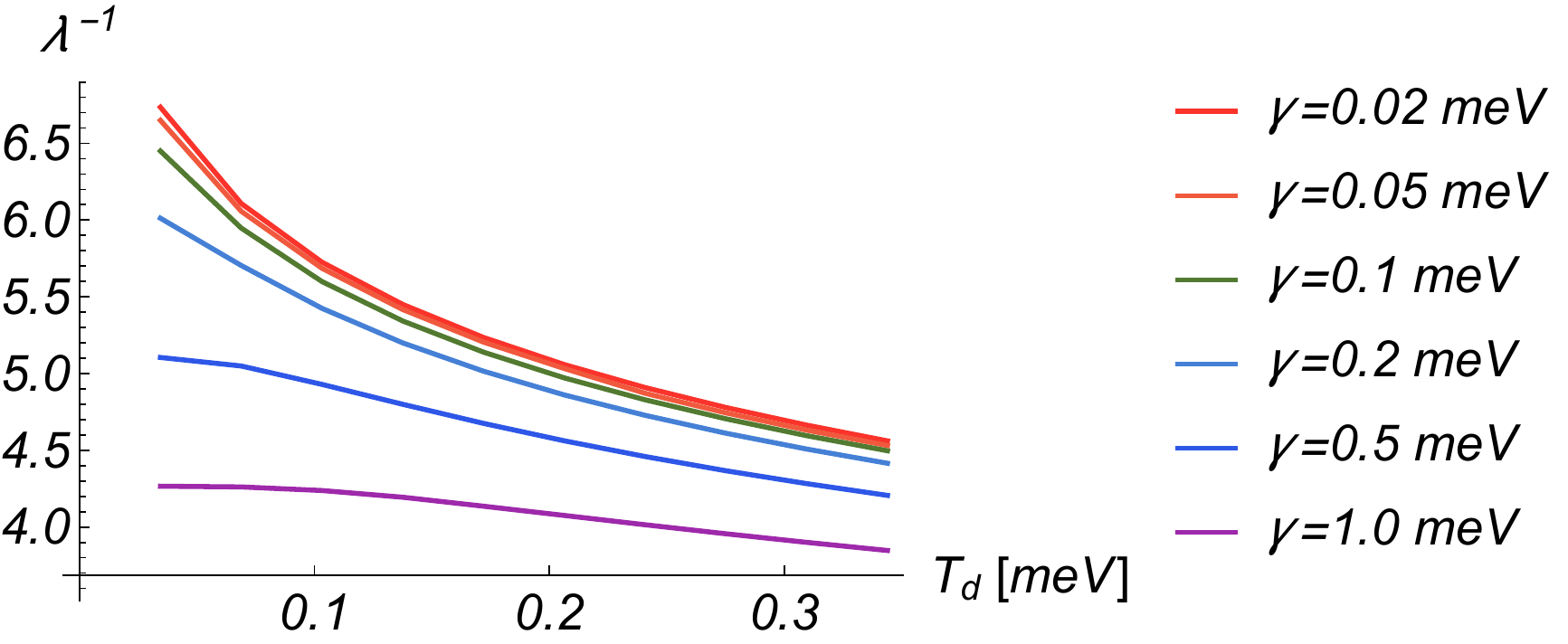}
    \caption{Relationship between $\lambda^{-1}_d$ and transition temperature $T_d$ evaluated including the proximity-induced self-energy. 
    Results are plotted for cutoff $\Lambda = 30\si{\meV}$ and substrate gap $\Delta_s = 1.057\si{\meV}$, corresponding to a substrate transition temperature of $T_s = 7\si{\kelvin}$, using $0^+ = .005\si{\meV}$ for a series of different tunneling size $\gamma$'s. }
    \label{fig:tc-fig-jon}
\end{figure}

This is complicated by the need to regularize the spectral functions with a factor of $0^+$, and also to cutoff the integrals over $\omega$ at a finite high-frequency cutoff, which we take to be $\omega^* = 100\si{\meV}$ (the integrands decay rapidly in frequencies above the cutoff $\Lambda$).
Throughout this we take $0^+ = .005\si{\meV}$.
All in all, we find the results including the proximity effect summarized in Fig.~\ref{fig:tc-fig-jon}.

\subsection{Collective Mode}
We now consider the fluctuation propagator by expanding around the saddle-point.
In the normal state with $\Delta^{cl} = 0$, such that we expand to quadratic order in $\check{\Delta}$. 
In addition to the Hubbard-Stratonovich term, which is already quadratic in $\Delta$, we also have the functional determinant.
Expanding this gives 
\begin{equation}
    \mathcal{S}_{\rm eff} = -\frac{1}{g_d}\sum_q \left( \overline{\Delta}^{cl}_q \Delta^{q}_q + \overline{\Delta}^{q}_q \Delta^{cl}_q  \right) +i\frac12 \Tr \check{\mathbb{G}}_0 \check{\Lambda}\check{\mathbb{G}}_0 \check{\Lambda} . 
\end{equation}
We can simplify the calculation by first invoking the fluctuation dissipation relation, such that we only need to calculate the retarded propagator, which is the classical-quantum component. 

Furthermore, we decompose the $d$-wave mode in to the real and imaginary parts with respect to the substrate gap (which we take to be real).
If we write 
\begin{equation}
    \Delta^{\alpha}_{q} = h_{q}^{\alpha} + i d^\alpha_{q}
\end{equation}
for $\alpha = q,cl$ we then can write the pairing vertex as 
\begin{equation}
   \tau^\dagger\check{\Delta}_{q} + \tau \check{\overline{\Delta}}_{-q} = \sigma_\alpha \left[ \tau_1 h_{q}^\alpha -\tau_2 d_q^{\alpha} \right],
\end{equation}
where we have introduced the short-hand notation that for $\alpha = q,cl$ we have $\sigma_q = \frac12\sigma_1$ and $\sigma_{cl} = \sigma_0$.

We expand the effective action in terms of the parameterized collective modes. 
The Hubbard-Stratonovich term has 
\begin{equation}
    S_{\rm HS} = -\frac{2}{g_d}\sum_q \left[h_{-q}^{cl}h^q_{q} + d^{cl}_{-q} d_{q}^{q} \right],
\end{equation}
and is therefore diagonal in this representation.
The quasiparticle contribution is 
\begin{equation}
     S_{\rm QP} = \frac{i}{2}\sum_{q}\int_p |\chi^d_{\bf p}|^2 \tr \left\{\check{\mathbb{G}}_{0}(p+\frac12 q) \sigma_\alpha \left[ \tau_1 h_{q}^\alpha -\tau_2 d_q^{\alpha} \right] \mathbb{\check{G}}_0(p-\frac12q)\sigma_\beta \left[ \tau_1 h_{-q}^\beta -\tau_2 d_{-q}^{\beta} \right] \right\} 
\end{equation}

Consider the cross-coupling between the $h$ and $d$ modes.
This involves a trace over the Green's functions and one vertex in the $\tau_1$ channel, with another in the $\tau_2$ channel. 
The only non-trivial contraction of the Nambu matrices must involve a $\tau_3$ in one Green's function and a $\tau_0$ in the other, and thus must be odd in $\xi$.
As such, in the quasiclassical limit this goes as $\int d\xi \xi$ and will nearly vanish due to approximate particle-hole symmetry (more accurately, it is small in $\Delta/E_F$). 
Therefore we neglect the cross coupling and see that the $h$ and $d$ modes decouple. 

The action for each is then found to be 
\begin{subequations}
\begin{align}
    & S_{dd} = \sum_q   \left[ -\frac{2}{g_d}d^{q}_{-q} d^{cl}_q+ d^{\alpha}_{q}d^{\beta}_{-q}\frac{i}{2}\int_p|\chi^d_{\bf p}|^2 \tr\mathbb{\check{G}}_0(p+\frac12q)\sigma_\alpha \tau_2 \mathbb{\check{G}}_0(p-\frac12 q )\sigma_\beta \tau_2  \right] \\
    & S_{hh} = \sum_q   \left[ -\frac{2}{g_d}h^{q}_{-q} h^{cl}_q+ h^{\alpha}_{q}h^{\beta}_{-q}\frac{i}{2}\int_p|\chi^d_{\bf p}|^2 \tr\mathbb{\check{G}}_0(p+\frac12q)\sigma_\alpha \tau_1 \mathbb{\check{G}}_0(p-\frac12 q )\sigma_\beta \tau_1  \right] .
\end{align}
\end{subequations}
Of these, the $q-q$ components are determined by fluctuation dissipation relation, so we focus on the $q-cl$ components.
We have 
\begin{subequations}
\begin{align}
    & S^{q-cl}_{dd} = \sum_q   \left[ -\frac{2}{g_d}d^{q}_{q} d^{cl}_{-q}+ d^{q}_{q}d^{cl}_{-q} i\int_p|\chi^d_{\bf p}|^2 \tr\mathbb{\check{G}}_0(p+\frac12q)\sigma_{q} \tau_2 \mathbb{\check{G}}_0(p-\frac12 q )\sigma_{cl} \tau_2 \right] \\
    & S^{q-cl}_{hh} = \sum_q   \left[ -\frac{2}{g_d}h^{q}_{q} h^{cl}_{-q}+ h^{q}_{q}h^{cl}_{-q}i\int_p|\chi^d_{\bf p}|^2 \tr\mathbb{\check{G}}_0(p+\frac12q)\sigma_{q} \tau_1 \mathbb{\check{G}}_0(p-\frac12 q )\sigma_{cl} \tau_1   \right].
\end{align}
\end{subequations}
This reduces down to the calculation of the correlation functions 
\begin{equation}
    (L^R)^{-1}_{ab}(q) = - \frac{2}{g_d}\delta_{ab} + i\int_p|\chi^d_{\bf p}|^2 \tr\mathbb{\check{G}}_0(p-\frac12q)\sigma_{q} \tau_{a} \mathbb{\check{G}}_0(p+\frac12 q )\sigma_{cl} \tau_{b}, 
\end{equation}
and in particular we have the $d$-mode propagator in the $a=b=2$ channel and the $h$ mode in the $a= b = 1$ channel. 

We evaluate the trace over the Keldysh matrices to find 
\begin{equation}
    (L^R)^{-1}_{ab}(q) = - \frac{2}{g_d}\delta_{ab} + \frac{i}{2}\int_p|\chi^d_{\bf p}|^2 \tr\left[  \tau_b\mathbb{\hat{G}}^K(p-\frac12q)\tau_a \mathbb{\hat{G}}^R(p+\frac12 q )  + \tau_a\mathbb{\hat{G}}^K(p+\frac12q)\tau_b \mathbb{\hat{G}}^A(p-\frac12 q ) \right].
\end{equation}

We evaluate this at center of mass momentum $\mathbf{q}= 0$, to simplify our analysis. 
In this case, the quasiclassical approximation may be invoked and we can average over the Fermi surface.
This removes the $d$-wave form factors and gives 
\begin{equation}
    \frac{1}{\nu_F}(L^R)^{-1}_{ab}(\Omega,\mathbf{q} = 0) = - 2\lambda^{-1}_d\delta_{ab} + \frac{i}{2}\int\frac{d\varepsilon}{2\pi} \int d\xi \tr\left[  \tau_b\mathbb{\hat{G}}^K(\varepsilon - \frac{\Omega}{2}, \xi )\tau_a \mathbb{\hat{G}}^R(\varepsilon + \frac{\Omega}{2},\xi )  + \tau_a\mathbb{\hat{G}}^K(\varepsilon + \frac{\Omega}{2},\xi)\tau_b \mathbb{\hat{G}}^A(\varepsilon - \frac{\Omega}{2}, \xi ) \right].
\end{equation}
We can simplify slightly by shifting $\varepsilon$ to get for the diagonal components
\begin{equation}
    \frac{1}{\nu_F}(L^R)^{-1}_{aa}(\Omega,\mathbf{q} = 0) = - 2\lambda^{-1}_d  + \frac{i}{2}\int\frac{d\varepsilon}{2\pi} \int d\xi \tau_a\mathbb{\hat{G}}^K(\varepsilon, \xi )\tau_a \left[ \mathbb{\hat{G}}^R(\varepsilon + \Omega ,\xi )  + \mathbb{\hat{G}}^A(\varepsilon - \Omega , \xi )\right] .
\end{equation}
In equilibrium we have 
\begin{equation}
    \mathbb{\hat{G}}^{K}(\varepsilon,\xi) =\tanh(\frac{\beta \varepsilon}{2}) \left[ \mathbb{\hat{G}}^R(\varepsilon,\xi) - \mathbb{\hat{G}}^A(\varepsilon,\xi) \right].
\end{equation}
In order to accelerate integrals and improve convergence we fold the integration over $-\varepsilon$ and integrate only over positive $\xi$.
The integrals over positive $\xi$ only are permissible because all terms odd in $\xi$ will not enter and therefore don't need to be cancelled. 
All terms in the given expression are either in the $\tau_0,\tau_3,\tau_1$ channels.
The Green's function trace will therefore involve traces of the terms $\tau_0^2,\tau_1^2,\tau_3^2,\tau_1,\tau_3,i\tau_2$.
Of these, the last three will vanish and only the first three survive, which are the squares of each individual term and therefore the trace will kill all terms odd in $\xi$. 

Thus, we evaluate this numerically as 
\begin{multline}
    \frac{1}{\nu_F}(L^R)^{-1}_{aa}(\Omega,\mathbf{q} = 0) = - 2\lambda^{-1}_d \\
    - \frac{1}{2i}\frac{1}{2\pi}\int_0^\infty d\varepsilon 2 \int_0^\Lambda d\xi \left( \tau_a\mathbb{\hat{G}}^K(\varepsilon, \xi )\tau_a \left[ \mathbb{\hat{G}}^R(\varepsilon + \Omega ,\xi )  + \mathbb{\hat{G}}^A(\varepsilon - \Omega , \xi )\right] +\tau_a\mathbb{\hat{G}}^K(-\varepsilon, \xi )\tau_a \left[ \mathbb{\hat{G}}^R(-\varepsilon + \Omega ,\xi )  + \mathbb{\hat{G}}^A(-\varepsilon - \Omega , \xi )\right] \right).
\end{multline}
In particular, we plot the spectral functions 
\begin{equation}
    \mathcal{A}_{dd}(\Omega) = -\frac{1}{\pi}\Im L^R_{22}(\Omega,\mathbf{q}=0)
\end{equation}
which are used to locate the collective mode resonances. 

\section{Dependence on Substrate Gap}

\begin{figure*}
    \centering
    \includegraphics[width=\linewidth]{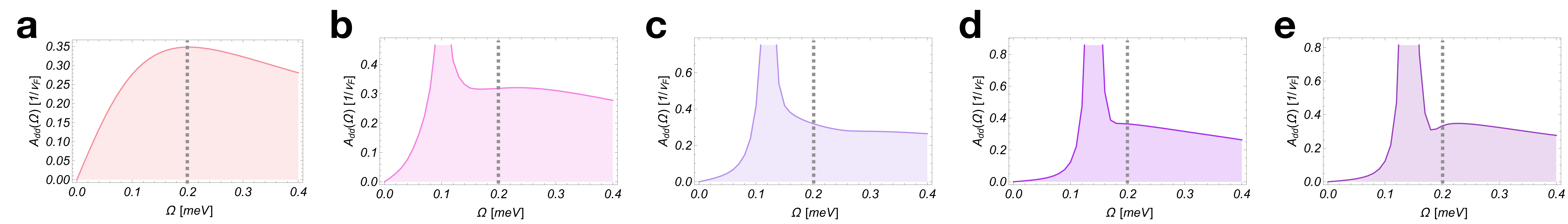}
    \caption{Collective mode spectral function for different values of $\Delta_s$ at fixed $\gamma = .2$meV and fixed $T = T_d^{(0)} = .344$meV and $\Lambda = 30$meV. 
    We plot the spectral function for $\Delta_s = 0,.1,.2,.5,1.0$meV in (a)-(e) respectively. 
    We see the evolution from an overdamped superconducting fluctuation in (a) in to a sharp collective mode in (e).
    Dashed line indicates the value of $\gamma$ in each plot, which is fixed at $\gamma = .2$meV. }
    \label{fig:bs-delta}
\end{figure*}

Here we briefly detail the dependence on the substrate gap, shown in Fig.~\ref{fig:bs-delta}

We see in particular that for finite $\gamma$ with $\Delta_s = 0 $ the substrate acts as a reservoir and broadens the electronic spectral function.
We then see no sharp mode in the $d$-wave channel, but instead it is replaced by an overdamped superconducting fluctuation.
In the absence of strong substrate effects this mode will have a lifetime which scales as $\tau^{-1} \sim T - T_d$, as it condenses at $T = T_d$~\cite{larkin-book}.
By inducing a substrate gap we crossover from the fluctuation regime, with Azlamazov-Larkin-type corrections to features in to the sharp collective mode outlined in the main text. 



\end{document}